\documentclass[12pt]{iopart}

\usepackage{graphicx}
\usepackage{bm}

\begin{document}

\title{Coexistence of opposite opinions in a network with communities.}

\author{ R. Lambiotte$^1$ and M. Ausloos$^{1}$}

\address{
$^1$ GRAPES, Universit\'e de Li\`ege, Sart-Tilman, B-4000 Li\`ege, Belgium
}
\ead{renaud.lambiotte@ulg.ac.be}

\begin{abstract}
The Majority Rule is applied to a topology that consists of two coupled random networks, thereby mimicking the modular structure observed in social networks.  We calculate analytically the asymptotic behaviour of the model and derive a phase diagram that depends on the frequency of random opinion flips and on the inter-connectivity between the two communities. It is shown that three regimes may take place: a disordered regime, where no collective phenomena takes place; a symmetric regime, where the nodes in both communities reach the same average opinion; an asymmetric regime, where the nodes in each community reach an opposite average opinion. The transition from the asymmetric regime to the symmetric regime is shown to be discontinuous.
\end{abstract}


\noindent{\it Keywords}: Random graphs, networks; Critical phenomena of socio-economic systems; Socio-economic networks

\maketitle

\section{Introduction}

In the last few years, the study of networks has received an enormous amount of attention from the scientific community \cite{review1,review2}, in disciplines as diverse as computer and information sciences (the Internet and the World Wide Web), sociology and epidemiology (networks of personal or social contacts), biology (metabolic and protein interaction), etc. This outburst of interest has been driven mainly by the possibility to use networks in order to represent 
many complex systems and by the availability of communication networks and computers that allow us to gather and analyze data on a scale far larger than previously possible. The resulting change of scale (from a few dozen of nodes in earlier works to several thousands of nodes today) has not only lead to the definition of new {\em statistical} quantities in order to describe {\em large} networks, e.g. degree distribution or clustering coefficient, but it has also addressed problems pertaining to Statistical Physics, for instance by looking at the interplay between the {\em microscopic} interactions of neighbouring nodes and  the behaviour of the system at a {\em macroscopic} level. 
Such a problem takes place in social networks,  i.e. nodes represent individuals and links between nodes represent their relations (e.g. friendship, co-authorship), when one tries to find macroscopic equations for the evolution of "society". Indeed, many studies have revealed  non-trivial structures in social networks, such as fat-tailed degree distributions \cite{bara}, a high clustering coefficient \cite{watts} and the presence of communities \cite{modular}.  A primordial problem is therefore to understand how this underlying topology influences the way the interactions between individuals (physical contact, discussions) may (or not) lead to collective phenomena. Typical examples would be the propagation of a virus \cite{boguna} or opinion \cite{galam,sznajd} in a social network, that may lead to the outbreak of an epidemics or of a new trend/fashion. 

It is now well-known that degree heterogeneity \cite{boguna,sood} is an important factor that may radically alter the macroscopic behaviour of a network but, surprisingly, the role played by its modular structure is still poorly known \cite{lambiotte,holyst}. It has been observed, though, that many social networks exhibit modular structures \cite{modular,modular2,modular3}, i.e. they are composed of highly connected communities, while nodes in different communities are sparsely connected. This lack of interaction between communities certainly has consequences on the way information diffuses through the network, for instance, but it also suggests that nodes belonging to different communities may behave in a radically different way.  

In this paper, we address such a problem by focusing on a simple model for networks with two communities, the Coupled Random Networks (CRN). To do so, one consider a set of $N$ nodes that one divides into two classes and one randomly assigns links between the nodes. Moreover, one assumes that the probability for two nodes to be linked is larger when they belong to the same class. Let us stress that CRN  has been first introduced in \cite{modular} and that it is particularly suitable in order to highlight the role of the network modularity while preserving its randomness. The {\em microscopic} dynamics that we apply on CRN is the well-known Majority Rule (MR) \cite{redner}. MR is a very general model for opinion-formation, i.e. nodes copy the behaviour of their neighbour, thereby suggesting that the results derived in this paper should also apply to other models of the same family. The effect of the inter-connectivity $\nu$ and of the frequency of random {\em flips}, measured by the parameter $q$ ($\sim temperature$) on the phase diagram is studied analytically. It is shown that three regimes may take place, depending on the parameters and on the initial conditions:  a disordered regime, where no collective phenomena takes place; a symmetric regime, where the nodes in both communities reach the same average opinion; an asymmetric regime, where the nodes in each community reach an opposite average opinion. The transition from the asymmetric regime to the symmetric regime is shown to be discontinuous. It is remarkable to note that a similar discontinuous transition also takes place when one applies MR to another network with communities, namely the Coupled Fully-Connected Networks introduced in \cite{lambiotte}. The main advantage of CRN is that its simpler structure allows to perform all the calculations exactly and to consider the case of a non-vanishing $q$ in detail. 

\section{Majority Rule}

The network is composed of $N$ nodes, each of them endowed with an opinion that can be either $\alpha$ or $\beta$. At each time step, one of the nodes is randomly selected and two processes may take place.  With probability $q$, the selected node  randomly picks an opinion $\alpha$ or $\beta$, whatever its previous opinion or the opinion of its neighbours. 
 With probability $1-q$, two neighbouring nodes of the selected node are also selected and the three agents in this {\em majority triplet} all adopt the state of the local majority (see Fig.1). The parameter $q$ therefore measures the competition between individual choices, that have a tendency to randomize the opinions in the system, and neighbouring interactions, that tend to homogenize the opinion of agents. In the case $q=0$, it is well-known that the system asymptotically reaches global consensus where all nodes share the same opinion \cite{redner}. In the other limiting case $q=1$, the system is purely random and the average (over the realizations of the random process) number of nodes with opinion $\alpha$ at time $t$, denoted by $A_t$, goes to $N/2$. 
 
 \begin{figure}
\includegraphics[width=6.5in]{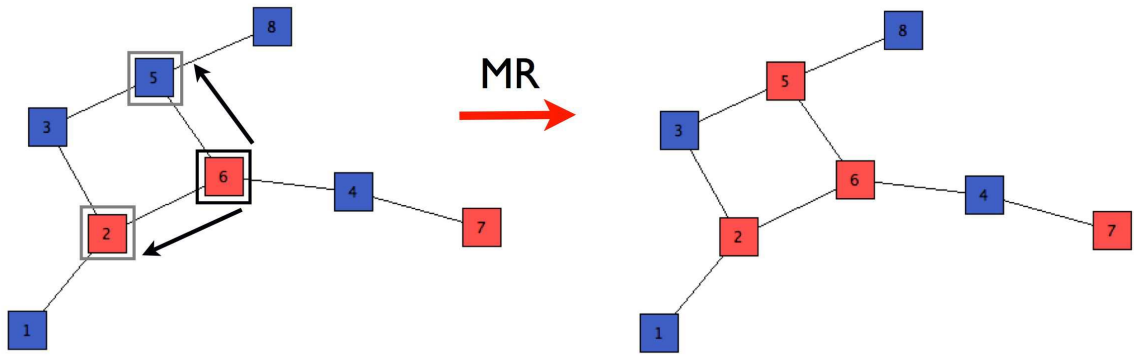}
\caption{Sketch of a time step, where one node (surrounded in black) and two of its neighbours (surrounded in grey) are selected. The majority rule implies that the blue node become red.}
\label{fig1}
\end{figure}

Let us first focus on a  network of individuals that are highly connected (in order to justify the use of mean-field methods) and where all the nodes are equivalent.  That case has been studied in detail elsewhere \cite{lam0}, and is repeated here for the sake of clarity and introducing notations. It is straightforward to show that the mean-field rate equation for $A_t$ reads 
\begin{eqnarray}
\label{simple}
A_{t+1} = A_t +  q (\frac{1}{2} - a_t ) - 3 (1-q) a_t (1 - 3 a_t + 2 a^2_t),
\end{eqnarray}
where $a_t=A_t/N$ is the average proportion of nodes with opinion $\alpha$.
The term proportional to $q$ accounts for the individual random flips. The second term, accounting for majority processes, is evaluated by calculating the probability that the majority triplet is composed of two nodes $\alpha$  and one node $\beta$,  $3 a_t^2 (1-a_t)$, or of two nodes $\beta$ and one node $\alpha$,  $3 a_t (1-a_t)^2$. Consequently, the total contribution to the evolution of $A_t$  is
\begin{eqnarray}
\label{w}
3 \left( a_t^2 (1-a_t)-a_t (1-a_t)^2 \right) = - 3 a_t (1 - 3 a_t + 2 a_t^2).
 \end{eqnarray} 
It is easy to show that $a=1/2$ is always a stationary solution of Eq.(\ref{simple}), as expected from symmetry reasons. $a=1/2$ corresponds to a disordered state where no collective opinion has emerged in the system. It is useful to rewrite the evolution equation for the quantities $\Delta_t=A_t-N/2$ and $\delta_t=\Delta_t/N=a-1/2$
\begin{eqnarray}
\label{simpleSim}
\Delta_{t+1} = \Delta_t + \frac{\delta_t}{2} \left(3-5q - 12 (1-q) \delta_t^2 \right),
\end{eqnarray}
from which one finds that the disordered solution $a=1/2$ ceases to be stable when $q<3/5$. In that case, the system reaches one of the following asymmetric solutions
\begin{eqnarray}
\label{solution}
a_- &=& \frac{1}{2} - \sqrt{\frac{3 - 5 q}{12(1-q)}} \cr
a_+ &=& \frac{1}{2} +  \sqrt{\frac{3 - 5 q}{12(1-q)}}.
\end{eqnarray}
The system therefore undergoes an order-disorder transition at $q=3/5$. Under this value, a collective opinion has emerged due to the {\em imitation} between neighbouring nodes. 
 In the limit case $q \rightarrow 0$, one finds $a_-=0$ and $a_+=1$ in agreement with the results of \cite{redner}.

\section{Role of communities}

\subsection{Coupled Random Networks}

Distinct communities within networks are defined as subsets of nodes which are more densely linked when compared to the rest of the network. For the sake of simplicity, we restrict the scope to networks composed of only two communities, denoted by 1 and 2. Our goal is to build an uncorrelated random network where nodes in $1$ are more likely to be connected with nodes in $1$ than with nodes in $2$, and vice-versa. To do so, we consider a set of $N$ nodes that we divide nodes into two classes, $1$ and $2$.  We evaluate each pair of nodes in the system and draw a link between these nodes with probability $p_{ij}$, where $i$ and $j$ $\in\{1,2\}$ are the class to which the two nodes belong. In the following, we consider a case where the number of nodes in 1 and 2, respectively denoted by $N_{1}$ and $N_{2}$, are equal $N_{1}=N_{2}=N/2$. Moreover, we restrict the scope to the following probabilities  $p_{12}=p_{21}=p_{cross}$ and $p_{11}=p_{22}=p_{in}$. By construction, nodes in $1$ are therefore connected on average to $k_{in}=p_{in} (N-1)/2\approx p_{in} N/2$ nodes in 1 and to $k_{cross} = p_{cross} N/2$ nodes in $2$, while nodes in $2$ are connected to $k_{cross}$ nodes in 1 and $k_{in}$ nodes in $2$. 
Let us stress that this approach pertains to the theory of networks with {\em hidden variables} \cite{boguna2,fron} where the probability for a pair to be linked would not factorize $p_{ij}\neq p_i p_j$. A similar model has already been used in order to test methods for community detection in \cite{modular}.

 \begin{figure}
\includegraphics[width=6.2in]{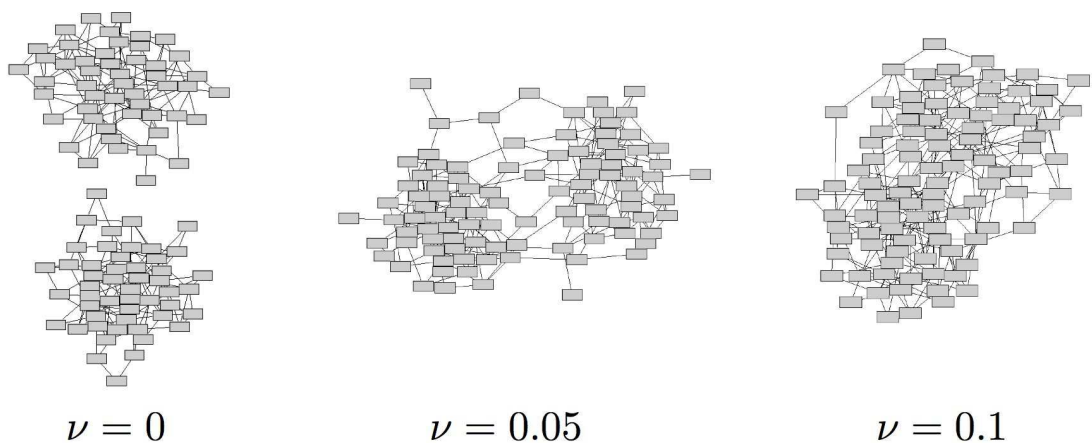}
\caption{Typical realizations of coupled random networks for small values of $\nu$. The network is composed of $N=100$ nodes and $p_{in}=0.1$. The system obviously consists of two communities that are less and less discernible for larger values of $\nu$. The graphs were plotted thanks to the {\em visone} graphical tools \cite{visone}.}
\label{fig2}
\end{figure}
 
This binary mixture, that we call a {\em Coupled Random Network} (CRN) is particularly suitable in order to reveal the role of network modularity \cite{modular}. Indeed, the inter-connectivity between the communities is tunable through the parameter $\nu=p_{cross}/p_{in}$. In the following, we focus on the interval $\nu \in [0,1]$ for which inter-community links are less frequent than intra-community links. When $\nu \rightarrow 1$, one recovers an homogeneous network where all nodes are {\em a priori} equivalent, while when $\nu<<1$, the communities are sparsely connected with each other.

Before going further, one should also point to an alternative model of modular networks introduced in \cite{lambiotte}. In that paper, Coupled Fully Connected Networks are composed of three kinds of nodes, the interface nodes, the nodes of type 1 and the nodes of type 2, and an inter-connectivity parameter measures the modularity of the network. MR has been applied on this topology with  $q=0$ and it has been shown that a discontinuous transition from an asymmetric to a symmetric state takes place. In the following, we will not only show that  a similar discontinuous transition takes place on CRN,  but we will also study analytically the behaviour of the system for $q \neq 0$

\subsection{Equation of evolution}

Let us denote by $A_{1}$ and $A_{2}$ the average number of nodes with opinion $\alpha$ among the two types of nodes. Let us first focus on the contributions when majority triplets are selected. At each time step, the probability that the selected node belongs to the first community is $1/2$. In that case, the probability that a randomly chosen link around the selected node goes to a node in $1$ is $k_{in}/(k_{in}+k_{cross})=1/(1+\nu)$. The probability that this randomly chosen link goes to a node in $2$ is $k_{cross}/(k_{in}+k_{cross})=\nu/(1+\nu)$. 
Consequently, the probability that the selected node belongs to $1$ and that both of its selected neighbours belong to $2$ is
\begin{eqnarray}
 \frac{1}{2} \frac{\nu^2}{(1+\nu)^2}.
\end{eqnarray}
Similarly, the probability that the selected node belongs to $1$, that one of its neighbours belongs to $1$ and that the other neighbour belongs to $2$ is
\begin{eqnarray}
\frac{1}{2} \frac{2 \nu}{(1+\nu)^2},
\end{eqnarray}
while the probability that all three nodes belong to $1$ is
\begin{eqnarray}
 \frac{1}{2} \frac{1}{(1+\nu)^2}.
\end{eqnarray}
The probabilities of events when the selected node belongs to $2$ are found in the same way. Putting all contributions together, one finds the probabilities $P_{(x,y)}$ that $x$ nodes 1 and $y$ nodes 2 belong to the majority triplet
\begin{eqnarray}
P_{(3,0)} &=&   \frac{1}{2} \frac{1}{(1+\nu)^2}\cr
P_{(2,1)} &=&   \frac{1}{2} \frac{2 \nu}{(1+\nu)^2} + \frac{1}{2} \frac{\nu^2}{(1+\nu)^2} = \frac{1}{2} \frac{\nu^2 + 2 \nu}{(1+\nu)^2} \cr
P_{(1,2)} &=& \frac{1}{2} \frac{\nu^2 + 2 \nu}{(1+\nu)^2} \cr
P_{(0,3)} &=& \frac{1}{2} \frac{1}{(1+\nu)^2}
\end{eqnarray}
where the normalization $\sum_{xy} P_{(x,y)}=1$ is verified. In order to derive coupled equations  for $A_{1;t}$ and $A_{2;t}$ that would generalize Eq.(\ref{simple}), one needs to evaluate the evolution of these quantities when a triplet $(x,y)$ is selected. To do so, one follows the steps described in \cite{lambiotte} and, when $q=0$, one obtains the equation of evolution
 \begin{eqnarray}
 \label{complicated}
  A_{1;t+1} - A_{1;t}&=&   \frac{3}{2} \frac{1}{(1+\nu)^2} (a_1^2 b_1- a_1 b_1^2)+  \frac{\nu^2 + 2 \nu}{(1+\nu)^2} (a_2 a_1 b_1- a_1 b_2 b_1)\cr
  &+&  \frac{1}{2} \frac{\nu^2 + 2 \nu}{(1+\nu)^2} (a^2_2 b_1- a_1 b^2_2) \cr
   A_{2;t+1} - A_{2;t}&=& \frac{3}{2} \frac{1}{(1+\nu)^2} (a_2^2 b_2- a_2 b_2^2)+ \frac{\nu^2 + 2 \nu}{(1+\nu)^2} (a_1 a_2 b_2- a_2 b_1 b_2)\cr
   &+&  \frac{1}{2} \frac{\nu^2 + 2 \nu}{(1+\nu)^2} (a^2_1 b_2- a_2 b^2_1) ,
 \end{eqnarray} 
  where $a_i$ and $b_i$ are respectively the proportion of nodes with opinion $\alpha$ and $\beta$  in the community $i$. After incorporating the term due to random flips, proportional to $q$, one obtains the set of non-linear equations
  
   \begin{eqnarray}
 \label{complicated2}
  A_{1;t+1} - A_{1;t}&=& \frac{q}{4} - \frac{q a_1}{2} + (1-q) [ \frac{3}{2} \frac{1}{(1+\nu)^2} (a_1^2 b_1- a_1 b_1^2) \cr
  &+&  \frac{\nu^2 + 2 \nu}{(1+\nu)^2} (a_2 a_1 b_1- a_1 b_2 b_1)
  +  \frac{1}{2} \frac{\nu^2 + 2 \nu}{(1+\nu)^2} (a^2_2 b_1- a_1 b^2_2)] \cr
   A_{2;t+1} - A_{2;t}&=& \frac{q}{4} - \frac{q a_2}{2} + (1-q) [ \frac{3}{2} \frac{1}{(1+\nu)^2} (a_2^2 b_2- a_2 b_2^2)\cr
   &+& \frac{\nu^2 + 2 \nu}{(1+\nu)^2} (a_1 a_2 b_2- a_2 b_1 b_2)
   +  \frac{1}{2} \frac{\nu^2 + 2 \nu}{(1+\nu)^2} (a^2_1 b_2- a_2 b^2_1)].
 \end{eqnarray} 
 Direct calculations show that Eq.(\ref{complicated2}) reduces to Eq.(\ref{simple}) in the limit $\nu=1$, as expected due to the indistinguishability of the nodes in that case.
 
\subsection{Stability of the disordered solution}

It is straightforward to show that $a_1=1/2$, $a_2=1/2$ is always a stationary solution of Eq.(\ref{complicated2}), whatever the values of $\nu$ and $q$. This solution consists of a disordered state where both communities behave similarly and where no favourite opinion has emerged due to MR.
 We study the stability of this disordered state by looking at small deviations $\epsilon_1=a_1-1/2$ and $\epsilon_2=a_2-1/2$ and keeping only linear corrections. In the continuous time limit and after re-scaling the time units, the evolution equations for these deviations read
 
    \begin{eqnarray}
 \label{linear1/2}
  \partial_t \epsilon_1 &=&- \frac{-3 + 5 q + 2 \nu (1+q) + \nu^2 (1+q)}{4 (1+\nu)^2} \epsilon_1 + \frac{\nu (2+\nu) (1-q)}{(1+\nu)^2} \epsilon_2 \cr
  \partial_t \epsilon_2 &=&  \frac{\nu (2+\nu) (1-q)}{(1+\nu)^2} \epsilon_1 - \frac{-3 + 5 q + 2 \nu (1+q) + \nu^2 (1+q)}{4 (1+\nu)^2} \epsilon_2.
 \end{eqnarray} 
 The eigenvalues of this linearized matrix of evolution are
 \begin{eqnarray}
 \lambda_1=  \frac{(3 - 10 \nu - 5 \nu^2 - 5 q + 6 \nu q + 3 \nu^2 q)}{4 (1 
+ \nu)^2}\cr
\lambda_2= \frac{1}{4}(3 - 5 q).
  \end{eqnarray} 
  By definition, the disordered solution is stable only when both eigenvalues are negative \cite{nicolis}, thereby ensuring that a small perturbation asymptotically vanishes. It is easy to show that only the values of $q$ in the interval $]3/5,1]$ respect this condition, whatever the value of $\nu$. This implies that the location of the order-disorder transition is not affected by the modularity of the network.  Let us also stress that that $\lambda_1$ goes to $\lambda_2$ when $\nu=0$. This is expected, as the system is composed of two independent networks in that case, so that the equations of evolution for $A_1$ and $A_2$ are decoupled. 
  
 \begin{figure}
\includegraphics[width=6.25in]{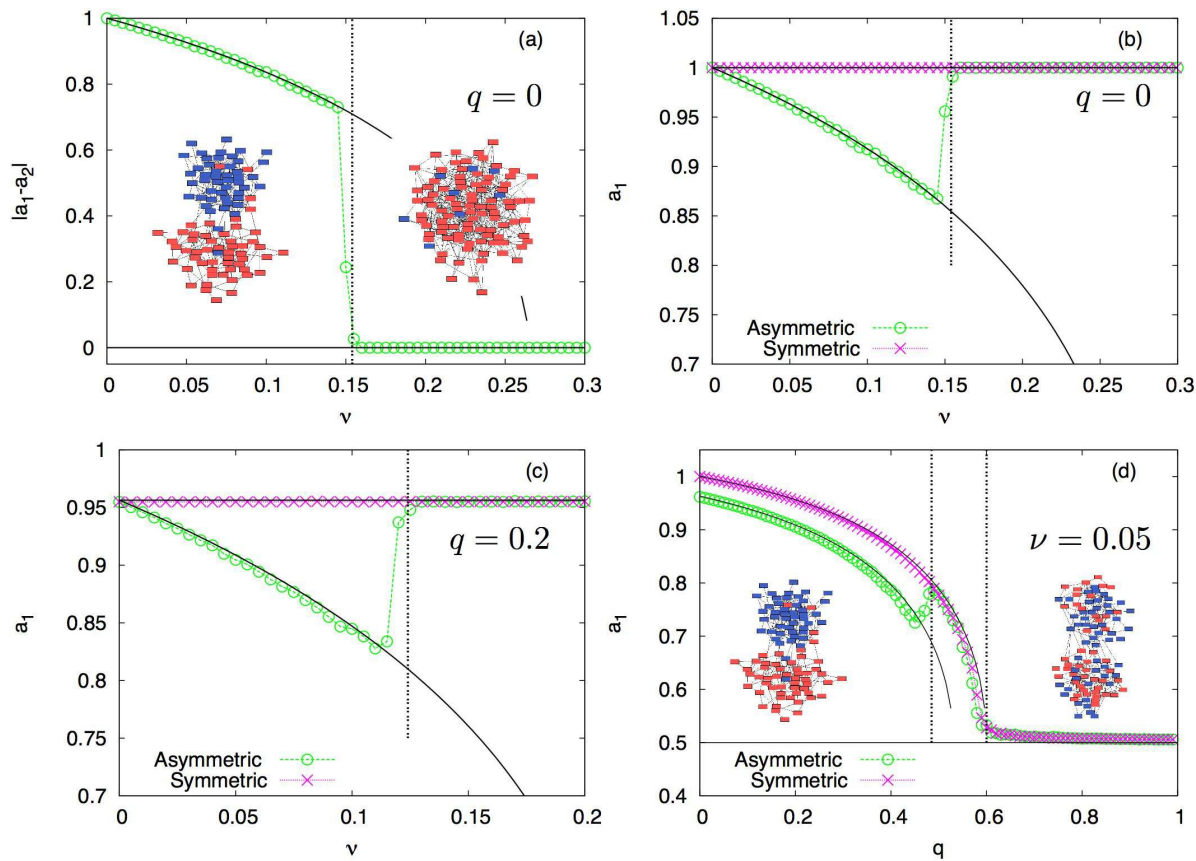}
\caption{Computer simulations of MR on Coupled Random Networks with  $N=10^4$ nodes and $p_{in}=0.01$. The simulations are stopped after $10^3$ steps/node and the results averaged over 100 realizations of the random process. The vertical dashed lines point to the theoretical transition point obtained from Eq.(\ref{final}) and to the critical value $q=3/5$ (Fig.3d). The solid lines correspond to the theoretical predictions (\ref{symsol}) and (\ref{solA}). The simulations are either started with a symmetric initial condition $a_1=1$, $a_2=1$ or with an asymmetric initial condition $a_1=1$, $a_2=0$. (a) Bifurcation diagram of $|a_1-a_2|$ as a function of $\nu$, for simulations starting from asymmetric initial conditions. The system ceases to be asymmetric $|a_1-a_2|>0$ above $\nu_c \approx 0.15$. (b) and (c) Bifurcation diagram of $a_1$ as a function of $\nu$, starting the simulations from asymmetric or symmetric initial conditions for $q=0$ (b) and $q=0.2$ (c). (d) Bifurcation diagram of $a_1$ as a function of $q$, starting the simulations from asymmetric or symmetric initial conditions for $\nu=0.05$. In that case, the system may undergo two transitions: one from the asymmetric to the symmetric state at $q \approx 0.485$, and one from the symmetric to the disordered state at $q=3/5$.}
\label{fig3}
\end{figure}
  
\subsection{Symmetric solution}

Our knowledge of the case $\nu=1$ (section 2) and the natural symmetry of CRN suggests to look at solutions of the form $a_1=1/2+\delta_S$, $a_2=1/2+\delta_S$ (S for symmetric). By inserting this ansatz into Eq.(\ref{complicated})
 \begin{eqnarray}
 \frac{q}{4} - \frac{q (\frac{1}{2}+\delta_S)}{2} + (1-q) [  \frac{3}{2} \frac{(\frac{1}{2}+\delta_S)^2 (\frac{1}{2}-\delta_S)- (\frac{1}{2}-\delta_S)^2 (\frac{1}{2}+\delta_S)}{(1+\nu)^2} \cr
+  \frac{\nu^2 + 2 \nu}{(1+\nu)^2} ((\frac{1}{2}+\delta_S)^2 (\frac{1}{2}-\delta_S)- (\frac{1}{2}-\delta_S)^2 (\frac{1}{2}+\delta_S))\cr
+  \frac{1}{2} \frac{\nu^2 + 2 \nu}{(1+\nu)^2} ((\frac{1}{2}+\delta_S)^2 (\frac{1}{2}-\delta_S)- (\frac{1}{2}-\delta_S)^2 (\frac{1}{2}+\delta_S))]=0 , 
\end{eqnarray} 
 direct calculations lead to the relation
  \begin{eqnarray}
  \label{condition}
- \frac{q \delta_S}{2} + (\frac{ \delta_S}{2}- 2 \delta_S^3) \frac{(1-q)}{(1+\nu)^2} [   \frac{3}{2}  +   \frac{3}{2}  \nu^2 + 3 \nu] = 0.
 \end{eqnarray} 
Let us insist on the fact that Eq.(\ref{condition}) is exact and not an expansion for small $\delta_S$.  It is a direct consequence of the above {\em symmetry} ansatz. The disordered solution $\delta_S=0$ obviously satisfies Eq.(\ref{condition}), but 
 symmetric solutions are also possible if they satisfy
 \begin{eqnarray}
- \frac{q (1+\nu)^2}{2 (1-q)}  + (\frac{ 1}{2}- 2 \delta_S^2)  [   \frac{3}{2}  +   \frac{3}{2}  \nu^2 + 3 \nu]=0,
 \end{eqnarray} 
 so that symmetric solutions have the form
  \begin{eqnarray}
 \label{symsol}
 \delta_S^2  = \frac{3 - 5 q}{12(1-q)}.
 \end{eqnarray} 
 It is remarkable to note that the symmetric solution does not depend on the inter-connectivity $\nu$. This is checked by comparing (\ref{symsol}) with the solution (\ref{solution}) obtained when the system is composed of only one community. It is also straightforward to show that (\ref{symsol}) is stable when $q<3/5$, as expected, from which one concludes that none of the characteristics of the order-disorder transition have been altered by the modularity of the network.  
    
\subsection{Asymmetric solution}
 
Let us first focus on the case $q=0$ where the dynamics is only driven by MR. We have shown above that the system may reach a symmetric frozen state $a_1=1$, $a_2=1$ or $a_1=0$, $a_2=0$ in that case (see Eq.(\ref{symsol})). However, computer simulations (Fig.3a) show that an asymmetric stationary state may prevail for small enough values of $\nu$. Computer simulations also show that the asymmetric state is characterized by averages of the form  $a_1=1/2+\delta_A$ and $a_2=1/2 - \delta_A$ (A for asymmetric).
Based on these numerical results, we look for stationary solutions of Eq.(\ref{complicated}) having this form. It is straightforward to show that the equations for $A_1$ and $A_2$ lead to the following condition
 \begin{eqnarray}
    (3 \delta_A- 12 \delta_A^3)+  (\nu^2 + 2 \nu) (- 2  \delta_A+ 8 \delta_A^3)
-  (\nu^2 + 2 \nu) (3 \delta_A + 4 \delta_A^3) =0,
 \end{eqnarray} 
 whose solutions are either $\delta_A=0$ (disordered state) or
 \begin{eqnarray}
 \label{solA}
\delta_A^2 = \frac{3- 5 (\nu^2 + 2 \nu) }{12  - 4 (\nu^2 + 2 \nu)}.
 \end{eqnarray} 
 The {\em disordered} solution $\delta_A=0$ is unstable when $q=0$, as shown above, and it it thus discarded. Let us also stress that the {\em asymmetric } solution differs from the frozen {\em symmetric} solution by the fact that fluctuations continue to take place in it. 
 
 By construction, the asymmetric solution exists only when $\delta_A^2 \geq 0$, namely when $\nu \in[0,\frac{-5+ \sqrt{40}}{5}\approx 0.26]$. In order to check the stability of (\ref{solA}) in this interval, we focus on the small deviations $ \epsilon_1 =a_1 - (1/2+\delta_A)$ and $\epsilon_2 = a_2-(1/2-\delta_A)$. After inserting these expressions into Eq.(\ref{complicated}) and keeping only linear terms, lengthy calculations lead to the eigenvalues
 \begin{eqnarray}
\lambda_1 = - \frac{3 - 10 \nu - 5 \nu^2}{2  (1+\nu)^2}\cr
\lambda_2 = - \frac{3}{2} \frac{1 - 6 \nu - 3 \nu^2}{  (1+\nu)^2},
 \end{eqnarray} 
from which one shows that the asymmetric solution loses its stability at a critical value
  \begin{eqnarray}
  \label{transi}
 \nu_c= ( \sqrt{48} -6)/6 \approx 0.15.
  \end{eqnarray} 
 Consequently, the system exhibits a discontinuous transition at $\nu_c$, as $1/2 + \delta_A(\nu_c)\approx0.85 \neq 1$. When $\nu<\nu_c$, the system may reach either the symmetric or the asymmetric solution depending on the initial conditions (and on the fluctuations). When $\nu>\nu_c$, in contrast, only the symmetric solution is attained in the long time limit. Let us stress that MR also undergoes a discontinuous transition from an asymmetric state to a symmetric state when it is applied to Coupled Fully Connected Networks \cite{lambiotte}. This similarity suggests therefore that such a transition is a generic feature of networks with modular structure. The above theoretical results have been verified by performing computer simulations of MR. The asymmetric solution (\ref{solA}) and the location of the transition (\ref{transi}) are  in perfect agreement with the simulations (Fig.3b).
 
  \begin{figure}
\includegraphics[width=6.in]{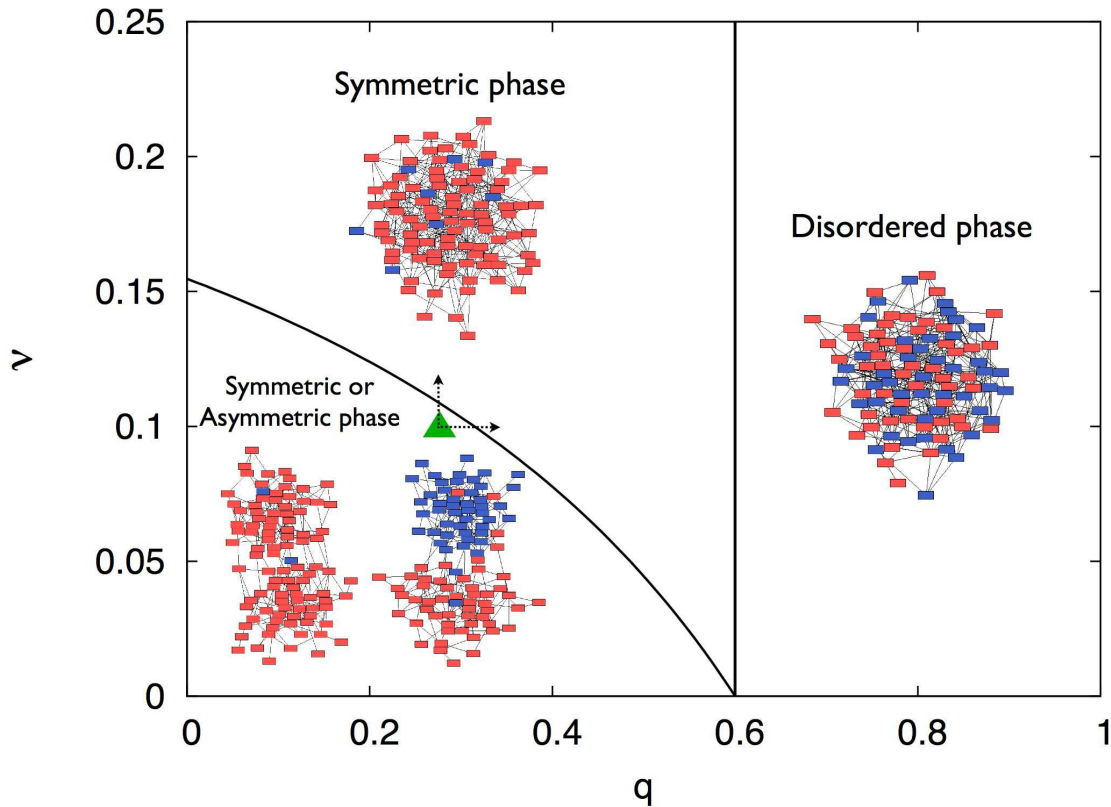}
\caption{Phase diagram of MR on CRN. Three phases may take place. i) a disordered phase when $q>3/5$; ii) a symmetric phase when $q<3/5$; iii) an asymmetric phase when $q<3/5$ and when $\nu< -1 + 2 \sqrt{\frac{ 3  q- 3 }{ 7 q - 9 }}$. A system in the asymmetric state, but  close to the transition line, e.g. the green triangle, may lose its stability due to an increase of the number of inter-community links (along $\nu$) or to an increase of the internal fluctuations (along $q$).}
\label{fig4}
\end{figure}
  
It is straightforward to generalize these results when $q> 0$. To do so, one inserts $a_1=1/2+\delta_A$, $a_2=1/2-\delta_A$ into Eq.(\ref{complicated2}) from which one obtains the relation
 \begin{eqnarray}
 \label{con2}
 - 2 q \delta_A + \frac{(1-q)}{(1+\nu)^2} [   (3 \delta_A- 12 \delta_A^3)+  (\nu^2 + 2 \nu) (- 2  \delta_A+ 8 \delta_A^3) \cr
-  (\nu^2 + 2 \nu) (3 \delta_A + 4 \delta_A^3) ] = 0.
 \end{eqnarray} 
 The stationary solutions of (\ref{con2}) are either $\delta_A=0$ or
 \begin{eqnarray}
 \label{sele}
\delta_A^2 = \frac{3 - 2 \frac{q (1+\nu)^2}{1-q} - 5 (\nu^2 + 2 \nu) }{12  - 4 (\nu^2 + 2 \nu)},
 \end{eqnarray} 
 and the  eigenvalues of the linearized equation of evolution around the stationary solution (\ref{sele}) are
\begin{eqnarray}
\lambda_1 = - \frac{3 - 10 \nu - 5 \nu^2 - 5 q + 6 \nu q + 3 \nu^2 q}{2  (1+\nu)^2}\cr
\lambda_2 = -  \frac{3 - 18 \nu - 9 \nu^2 - 5 q + 14 \nu q + 7 \nu^2 q}{2  (1+\nu)^2}.
 \end{eqnarray} 
The asymmetric solution is shown to loose its stability when
  \begin{eqnarray}
 3 - 18 \nu - 9 \nu^2 - 5 q + 14 \nu q + 7 \nu^2 q=0
   \end{eqnarray} 
 that one simplifies into
    \begin{eqnarray}
    \label{final}
\nu_c(q)= -1 + 2 \sqrt{\frac{ 3  q- 3 }{ 7 q - 9 }}.
  \end{eqnarray} 
This relation therefore determines the critical line above which only symmetric solutions prevail (see Fig.4). One can shown that  (\ref{final}) decreases with $q$ and that it goes to zero at the transition point $q=3/5$. It is also easy to show that the transition from the asymmetric to symmetric state is discontinuous for any values of $q<3/5$. Eq. (\ref{sele}) and Eq. (\ref{final}) have been successfully checked by computer simulations (see Fig.3c and Fig.3d).

\section{Discussions}

In this last section, we would like to point to some possible applications of this work and to the implications of our theoretical predictions. 
First of all, let us remind that many networks exhibit modular structures. It is therefore of significant practical importance to better understand the role played by these structures. This is true for social networks, where groups of nodes correspond to 
social communities or cliques, but also for the World Wide Web, where groups of highly connected sites may be related by a common topic \cite{info1,info2}, biological networks, where clusters of nodes may be associated to functional modules \cite{biology1,biology2,biology3},  GDP networks, from which dependencies between countries might be uncovered \cite{eco1,eco2} and even self-citation networks where clusters may reveal the field mobility of a scientist \cite{iina}. The identification of such communities is thus a very important problem, that has received a considerable amount of attention recently \cite{structure2,structure3,structure4,structure5,structure6,structure7,structure8,structure9}. Amongst others investigations, it has been shown that the use of  Ising-like models may be helpful in order to unravel communities in complex networks \cite{structurea,structure}.  This approach consists in applying a dynamics such as the majority rule on a pre-given network and to identify a community as a set of nodes having the same opinion/spin. In the language of this paper, the identification of the communities is possible only if the system reaches an asymmetric regime, i.e. if both communities reach different states. Our work could therefore provide a theoretical background for the use of such identification techniques. For instance, our results show that MR has a "minimum resolution", i.e. MR does not discriminate communities when $\nu>\nu_c(q)$ (with $\nu_c(q)<(\sqrt{48}-6)/6$)  because the  asymmetric state is not stable in that case. One expects that similar effects could also take place for similar models of opinion formation.

The fact that nodes in different communities exhibit different behaviours, i.e. the asymmetric regime, has been observed in many situations. Amongst others, one may think of  subcommunities in collaboration networks, that may correspond roughly to topics of research \cite{modular}, the Political Blogosphere, where it was observed that bloggers having a different political opinion are segregated \cite{adamic}, the existence of niche markets \cite{niche}, where small communities may use products that are different from those used by the majority, language dynamics, where it is well-known that natural frontiers may also coincide with a linguistic frontier \cite{language}, etc. The discontinuity of the transition from this asymmetric state to a symmetric "globalized" state might thus have radical consequences for systems close to the transition line: the addition of a few links between the communities or a small increase of the fluctuations inside the system (see Fig.4) may be sufficient in order to drive the system out of the asymmetric state. Such rapid shifts of behaviour of a whole sub-community should be searched in empirical data, in the dynamics of trends or fashion \cite{fashion} for instance.

To conclude, we would like to point to a possible generalization that would certainly make the model more realistic. In this paper, we have focused on a model of opinion formation evolving on a static topology. The effect of the network structure on the formation of a collective opinion has therefore been highlighted, but the fact that the opinion of the nodes themselves might influence the topology around them, e.g. links between nodes with different opinions could disappear, has not been considered. A model where the network inter-connectivity $\nu$ might co-evolve \cite{coe} with the node spins/opinions would thus be of high interest and will be considered in a future work. Other interesting generalizations would incorporate agents with ageing \cite{ageing0,klemm,lamA} or memory \cite{sornette}, the presence of leaders \cite{leaders}, etc.

 {\bf Acknowledgements}
This work has been supported by European Commission Project 
CREEN FP6-2003-NEST-Path-012864. RL thanks J. Ho{\l}yst, M. Buchanan and A. Scharnhorst for fruitful comments.

\end{document}